\documentclass[aps,prl,nofootinbib,twocolumn,superscriptaddress,letterpaper,amsmath,amssymb]{revtex4}

\usepackage{graphicx}
\usepackage{subfigure}
\def\missET {{\not\!\! E_T}}

\begin{document}

\title[Searching for Exotic Particles in High-Energy Physics with Deep Learning]{Searching for Exotic Particles in High-Energy Physics with Deep Learning}

\author{ P. Baldi}
\address{Dept. of Computer Science, UC Irvine, Irvine, CA 92617}
\email{pbaldi@uci.edu}
\author{ P. Sadowski}
\address{Dept. of Computer Science, UC Irvine, Irvine, CA 92617}
\author{D.~Whiteson}
\address{Dept. of Physics and Astronomy, UC Irvine, Irvine, CA 92617}
\email{daniel@uci.edu}
\begin{abstract}
Collisions at high-energy particle colliders are a traditionally fruitful source of exotic particle discoveries.  Finding these rare particles requires solving difficult signal-versus-background classification problems, hence machine learning approaches are often used. Standard approaches  have relied on `shallow' machine learning models that have a limited capacity to learn complex non-linear functions of the inputs, and rely on a pain-staking search through manually constructed non-linear features.  Progress on this problem has slowed, as a variety of techniques  have shown equivalent performance. Recent advances in the field of deep learning make it possible to learn more complex functions and better discriminate between signal and background classes.
Using benchmark datasets, we show that deep learning methods need no manually constructed inputs and yet improve the classification metric by as much as 8\% over the best current approaches. This demonstrates that deep learning approaches can improve the power of collider searches for exotic particles.
\end{abstract}

\maketitle
\section{Introduction}

The field of \emph{high energy physics} is devoted to the study of the
elementary constituents of matter.  By investigating the structure of
matter and the laws that govern its interactions, this field strives
to discover the fundamental properties of the physical universe.  The primary tools of experimental high energy physicists are modern
accelerators, which collide protons and/or antiprotons to create exotic
particles that occur only at extremely high energy densities. 
Observing these particles and measuring their properties may yield
critical insights about the very nature of matter~\cite{snowmasshiggs}.  Such discoveries require powerful statistical methods, and machine learning tools play a critical role. Given the limited quantity and expensive nature of the data, improvements in analytical tools directly boost particle discovery potential.

To discover a new particle, physicists must isolate a subspace of their high-dimensional data in which the hypothesis of a new particle or force gives a significantly different prediction than the null hypothesis, allowing for an effective statistical test. For this reason, the critical element of the search for new particles and forces in  high-energy physics is the computation of the {\it relative likelihood}, the ratio of the sample likelihood functions in the two considered hypotheses, shown by Neyman and Pearson~\cite{neymanpearson} to be the optimal discriminating quantity.
Often this relative likelihood function cannot be expressed analytically, so simulated collision data generated with Monte Carlo methods are used as a basis for approximation of the likelihood function.  The high dimensionality of data, referred to as the {\it feature space}, makes it intractable to generate enough simulated collisions to describe the relative likelihood in the full feature space, and machine learning tools are used for dimensionality reduction. Machine learning classifiers such as neural networks provide a powerful way to solve this learning problem.

The relative likelihood function is a complicated function in a high-dimensional space. While any function can theoretically be represented by a `shallow' classifier, such as a neural network with a single hidden layer \cite{hornik_multilayer_1989}, an intractable number of hidden units may be required. Circuit complexity theory tells us that deep neural networks (DN) have the potential to compute complex functions much more efficiently (fewer hidden units), but in practice they are notoriously difficult to train due to the vanishing gradient problem \cite{hochreiter_recurrent_1998,bengio1994learning}; the adjustments to the weights in the early layers of a deep network rapidly approach zero during training.
A common approach is to combine shallow classifiers with high-level features that are derived manually from the raw features. These are generally non-linear functions of the input features that capture physical insights about the data. While helpful, this approach is labor-intensive and not necessarily optimal; a robust machine learning method would obviate the need for this additional step and capture all of the available classification power directly from the raw data.

Recent successes in deep learning -- {\emph{e.g.}} neural networks with multiple hidden layers -- have come from alleviating the gradient diffusion problem by a combination of factors, including: 1) speeding up the stochastic gradient descent algorithm with graphics processors; 2) using much larger training sets; 3) using
new learning algorithms, including randomized algorithms such as dropout \cite{dropout2012,baldi_dropout_2014}; and 4)
pre-training the initial layers of the network with unsupervised learning methods such as autoencoders \cite{hinton_fast_2006, bengio_greedy_2007}. 
With these methods, it is becoming common to train deep networks of five or more layers. These advances in deep learning could have a significant impact on  applications in high-energy physics. Construction and operation of the particle accelerators is extremely expensive, so any additional classification power extracted from the collision data is very valuable.

In this paper, we show that the current techniques used in high-energy physics fail to capture all of the available information, even when boosted by manually-constructed physics-inspired features. This effectively reduces the power of the collider to discover new particles. We demonstrate that recent developments in deep learning tools can overcome these failings, providing significant boosts even without manual assistance.


\section{Results}

The vast majority of particle collisions do not produce exotic particles.  For example, though the Large Hadron Collider produces approximately $10^{11}$ collisions per hour, approximately 300 of these collisions result in a Higgs boson, on average.  Therefore, good data analysis depends on distinguishing collisions which produce particles of interest (\emph{signal})  from those producing other particles (\emph{background}).

Even when interesting particles are produced, detecting them poses considerable challenges.  They are too small to be directly observed and decay almost immediately into other particles. Though new particles cannot be directly observed, the lighter stable particles to which they decay, called
\emph{decay products}, can be observed.  Multiple layers of
detectors surround the point of collision for this purpose.
As each decay product pass through these detectors, it interacts with
them in a way that allows its direction and momentum to be measured. 

Observable decay products include electrically-charged leptons (electrons or muons, denoted $\ell$), and particle jets (collimated streams of particles originating from quarks or gluons, denoted $j$). In the case of jets we attempt to distinguish between jets from heavy quarks ($b$) and jets from gluons or low-mass quarks; jets consistent with $b$-quarks receive a  $b$-quark {\it tag}.  For each object, the momentum is determined by three measurements: the momentum transverse to the beam direction ($p_{\rm T}$), and two angles, $\theta$ (polar) and $\phi$ (azimuthal). For convenience, at hadron colliders, such as Tevatron and LHC, the {\it pseudorapidity}, defined
as $\eta = -\ln(\tan(\theta/2))$ is used instead of the polar angle $\theta$. Finally, an important quantity is the amount of momentum carried away by the invisible particles. This cannot be directly measured, but can be inferred in the plane transverse to the beam by requiring conservation of momentum.  The initial state has zero momentum transverse to the beam axis, therefore any imbalance of transverse momentum (denoted $\missET$) in the final state must be due to production of invisible particles such as neutrinos ($\nu$) or exotic particles. The momentum imbalance in the longitudinal direction along the beam cannot be measured at hadron colliders, as the initial state momentum of the quarks is not known.

\subsection*{Benchmark Case for Higgs Bosons (HIGGS)}

The first benchmark classification task is to distinguish between a signal process where new
theoretical  Higgs bosons are produced, and a background process with the
identical decay products but distinct kinematic features.  This benchmark task was recently considered by experiments at the LHC~\cite{atlaswwbb} and the Tevatron colliders~\cite{cdfwwbb}.

\begin{figure}
\centering

\subfigure[ ]{
\includegraphics[width=2in]{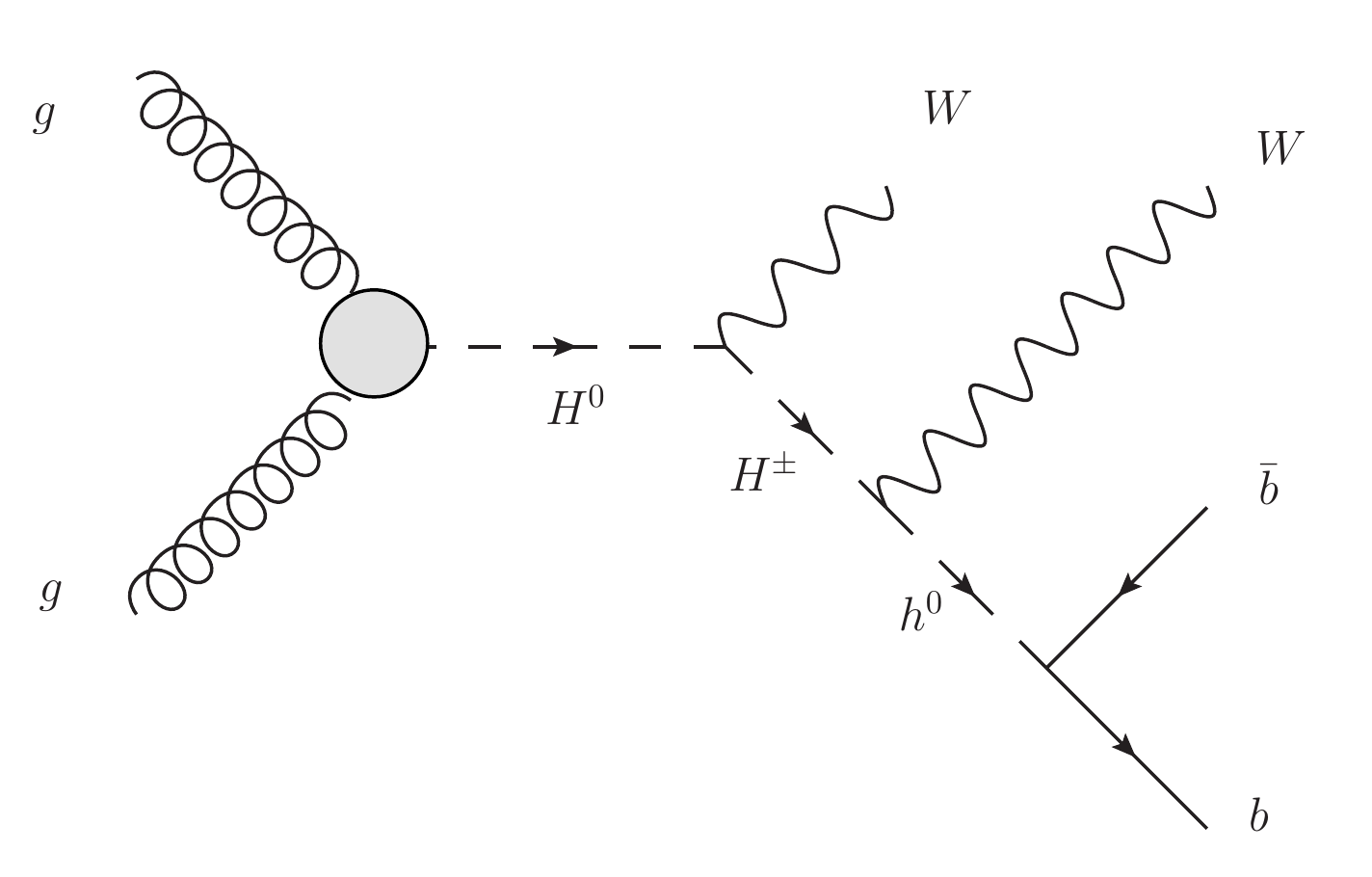}
\label{fig:ttbb_a}
}
\subfigure[ ]{
\includegraphics[width=2in]{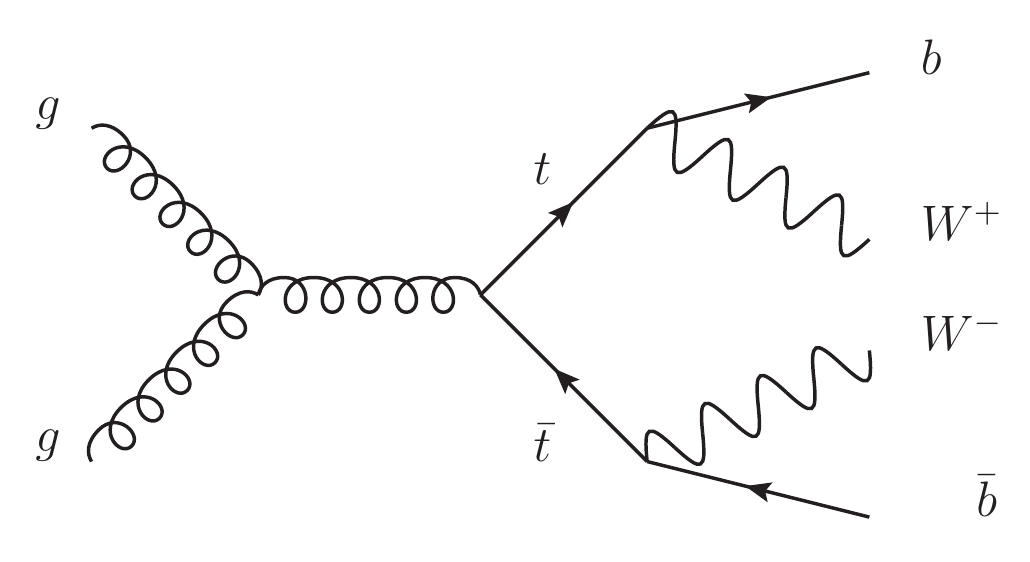}
\label{fig:ttbb_b}
}
\caption{{\bf Diagrams for Higgs benchmark}. (a) Diagram describing the signal process involving new exotic Higgs bosons $H^0$ and $H^\pm$. (b) Diagram describing the background process involving top-quarks ($t$). In both cases, the resulting particles are two $W$ bosons and two $b$-quarks.}
\label{fig:ttbb}
\end{figure}

The signal process is the fusion of two gluons into a heavy electrically-neutral
Higgs boson ($gg\rightarrow H^0$), which decays to a heavy electrically-charged Higgs bosons ($H^\pm$) and a $W$ boson. The $H^\pm$ boson subsequently decays to a second $W$ boson and the light Higgs boson, $h^0$ which has recently been observed by the ATLAS~\cite{atlashiggs}  and CMS~\cite{cmshiggs}  experiments. The light Higgs boson decays predominantly to a pair of bottom quarks, giving the process:
\begin{equation}
gg\rightarrow  H^0\rightarrow W^{\mp}H^{\pm}\rightarrow W^\mp W^\pm h^0\rightarrow
  W^\mp W^\pm b\bar{b}, 
\end{equation}

\noindent
which leads to $W^\mp W^\pm b\bar{b}$, see
Figure~\ref{fig:ttbb}.  The background process, which mimics  $W^\mp W^\pm b\bar{b}$  without the Higgs boson intermediate state, is the production of a pair of top quarks, each of which decay to $Wb$, also giving $W^\mp W^\pm b\bar{b}$, see Figure~\ref{fig:ttbb}.

Simulated events are generated with the {\sc madgraph}5~\cite{madgraph} event
generator assuming 8 TeV collisions of protons as at the latest run of the Large Hadron Collider, with showering and hadronization performed by
{\sc pythia}~\cite{SJO-0601} and detector response simulated by
{\sc delphes}~\cite{delphes}.  For the benchmark case here, $m_{H^0}=425$ GeV and $m_{H^\pm}=325$ GeV has been assumed.

We focus on the semi-leptonic decay mode, in which one $W$ boson
decays to a lepton and neutrino ($\ell\nu$) and the other decays to a pair of
jets ($jj$), giving decay products $\ell\nu b\ j j b$.  We consider events
which satisfy the requirements:

\begin{itemize}
\item Exactly one electron or muon, with $p_{T}>20$ GeV and
  $|\eta|<2.5$; 
\item at least four jets, each with $p_{T}>20$ GeV and
  $|\eta|<2.5$;
\item $b$-tags on at least two of the jets, indicating that they are likely due to $b$-quarks rather than gluons or lighter quarks.
\end{itemize}

\begin{figure}
\centering
\subfigure[ ]{
\includegraphics[width=1.5in]{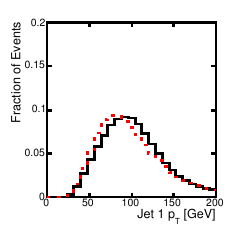}
\label{fig:llvar_a}}
\subfigure[ ]{
\includegraphics[width=1.5in]{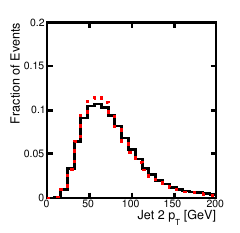}
\label{fig:llvar_b}}
\subfigure[ ]{
\includegraphics[width=1.5in]{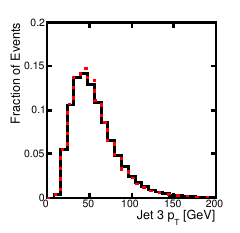}
\label{fig:llvar_c}}
\subfigure[ ]{
\includegraphics[width=1.5in]{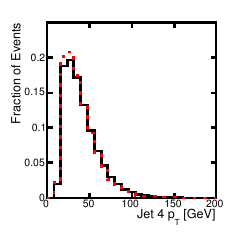}
\label{fig:llvar_d}}
\subfigure[ ]{
\includegraphics[width=1.5in]{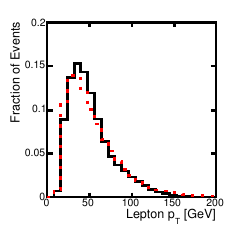}
\label{fig:llvar_e}}
\subfigure[ ]{
\includegraphics[width=1.5in]{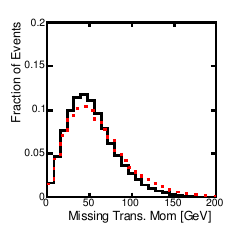}
\label{fig:llvar_f}}
\caption{ {\bf Low-level input features for Higgs benchmark.} Distributions in $\ell\nu jj b\bar{b}$ events for simulated signal (black) and background (red) benchmark events. Shown are the distributions of transverse momenta ($p_{\rm T}$) of each observed particle (a,b,c,d,e) as well as the imbalance of momentum in the final state (f). Momentum angular information for each observed particle is also available to the network, but is not shown, as the one-dimensional projections have little information. }
\label{fig:llvar}
\end{figure}

Events which satisfy the requirements above are naturally described by a simple set of features which represent the basic measurements made by the particle detector: the momentum of each observed particle. In addition we reconstruct the missing transverse momentum in the event and have $b$-tagging information for each jet. Together, these twenty-one features comprise our {\it low-level feature set}.  Figure~\ref{fig:llvar} shows the distribution of a subset of these  kinematic features for signal and background processes. 

The low-level features show some distinguishing characteristics, but our knowledge
of the different intermediate states of the two processes allows us to construct
other features which better capture the differences.  As the difference in the two hypotheses lies mostly in the existence of new intermediate Higgs boson states, we can distinguish between the two hypotheses by attempting to identify whether the intermediate state existed. This is done by reconstructing its characteristic invariant mass; if a particle A decays into particles B and C, the invariant mass of particle A ($m_A$) can be reconstructed as:

\begin{equation}
 m^2_{A} = m^2_{B+C} = (E_B + E_C)^2 - |({\mathbf{p}}_B+{\mathbf{p}}_C)|^2 
\end{equation}

\noindent where $E$ is the energy and $\mathbf{p}$ is the three-dimensional momentum of the particle.  Similarly, the invariant mass of more than two particles can be defined as the modulus of the particles' Lorentz four-vector sum. In the signal hypothesis we expect that:

\begin{itemize}
\item $W\rightarrow \ell\nu$ gives a peak in the $m_{\ell\nu}$ distribution at the known $W$ boson mass, $m_W$,
\item $W\rightarrow jj$ gives a peak in the $m_{jj}$ distribution at $m_W$,
\item $h^0\rightarrow b\bar{b}$ gives a peak in the $m_{b\bar{b}}$ distribution at the known Higgs boson mass, $m_{h^0}$,
\item $H^\pm\rightarrow Wh^0$ gives a peak in the $m_{Wb\bar{b}}$ distribution at the assumed $H^\pm$ mass, $m_{H^\pm}$,
\item $H^0\rightarrow WH^\pm$ gives a peak in the $m_{WWb\bar{b}}$ distribution at the assumed $H^0$ mass,at $m_{H^0}$,
\end{itemize}

Note that the leptonic $W$ boson is reconstructed by combining the lepton with neutrino, whose transverse momentum is deduced from the imbalance of momentum in the final state objects and whose rapidity is set to give $m_{\ell\nu}$ closest to $m_W=80.4$~GeV.

\noindent
while in the case of the $t\bar{t}$ background we expect that:

\begin{itemize}
\item $W\rightarrow \ell\nu$ gives a peak in $m_{\ell\nu}$ at $m_W$,
\item $W\rightarrow jj$ gives a peak in $m_{jj}$ at $m_W$,
\item $t\rightarrow Wb$ gives a peak in $m_{j\ell\nu}$ and $m_{jbb}$ at $m_t$.
\end{itemize}



\begin{figure}
\centering
\subfigure[ ]{
\includegraphics[width=1.5in]{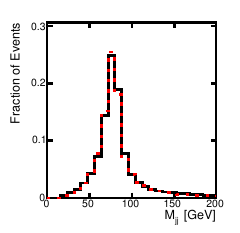}
\label{fig:hlvar_a}}
\subfigure[ ]{
\includegraphics[width=1.5in]{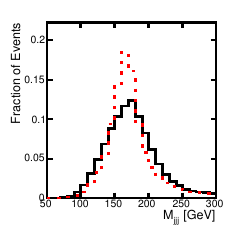}
\label{fig:hlvar_b}}
\subfigure[ ]{
\includegraphics[width=1.5in]{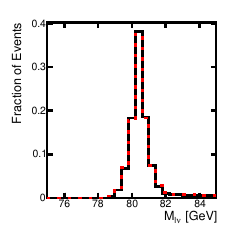}
\label{fig:hlvar_c}}
\subfigure[ ]{
\includegraphics[width=1.5in]{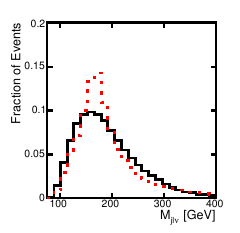}
\label{fig:hlvar_d}}
\subfigure[ ]{
\includegraphics[width=1.5in]{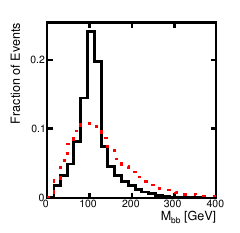}
\label{fig:hlvar_e}}
\subfigure[ ]{
\includegraphics[width=1.5in]{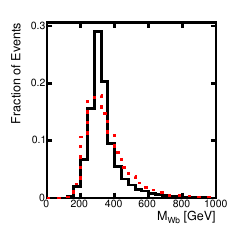}
\label{fig:hlvar_f}}
\subfigure[ ]{
\includegraphics[width=1.5in]{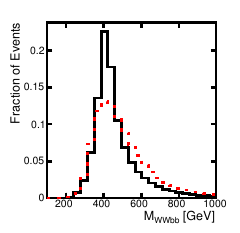}
\label{fig:hlvar_g}}
\caption{ {\bf High-level input features for Higgs benchmark.} Distributions in simulation of invariant mass calculations in $\ell\nu jj b\bar{b}$ events
  for simulated signal (black) and background (red) events.}
\label{fig:hlvar}
\end{figure}

See Figure~\ref{fig:hlvar} for distributions of these high-level
features for both signal and background processes. Clearly these contain more  discrimination power than the low-level features.

We have published a dataset containing 11 million simulated collision events for benchmarking machine learning classification algorithms on this task, which can be found in the UCI Machine Learning Repository at archive.ics.uci.edu/ml/datasets/HIGGS.

\subsection*{Benchmark Case for Supersymmetry Particles (SUSY)}

The second  benchmark classification task is to distinguish between a process where new supersymmetric particles are produced, leading to a final state in which some particles are detectable and others are invisible to the experimental apparatus, and a background process with the same detectable particles but fewer invisible particles and distinct kinematic features.  This benchmark problem is currently of great interest to the field of high-energy physics, and there is a vigorous effort in the literature~\cite{mt2,Barr:2003rg,razor,superrazor} to build high-level features which can aid in the classification task.

\begin{figure}
\centering
\subfigure[ ]{
\includegraphics[width=2in]{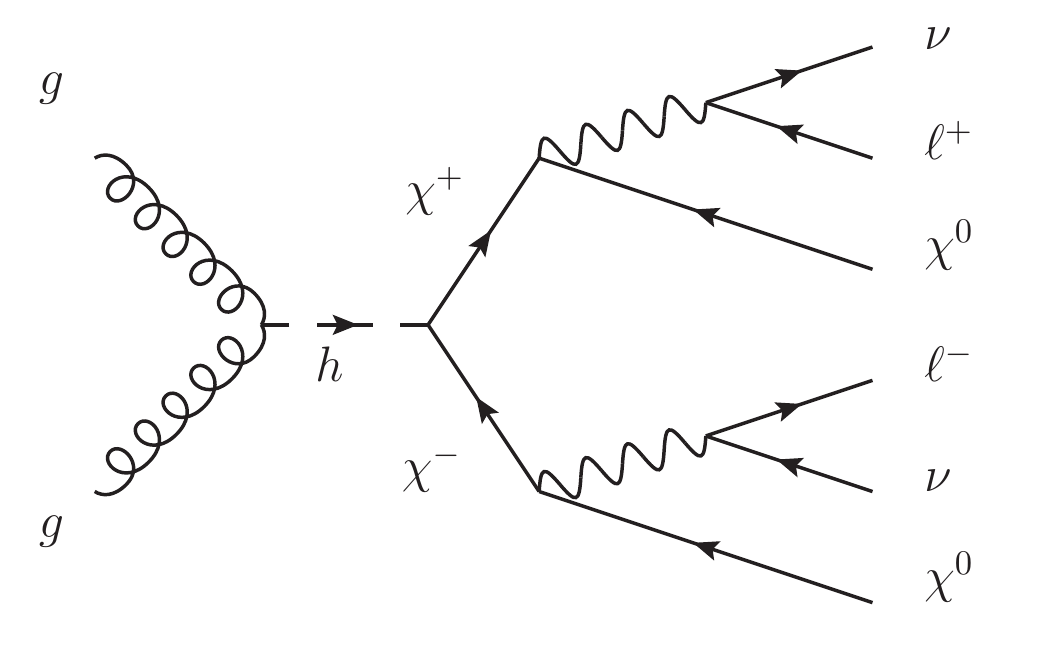}
\label{fig:xx_a}}
\hspace{1cm}
\subfigure[ ]{
\includegraphics[width=2in]{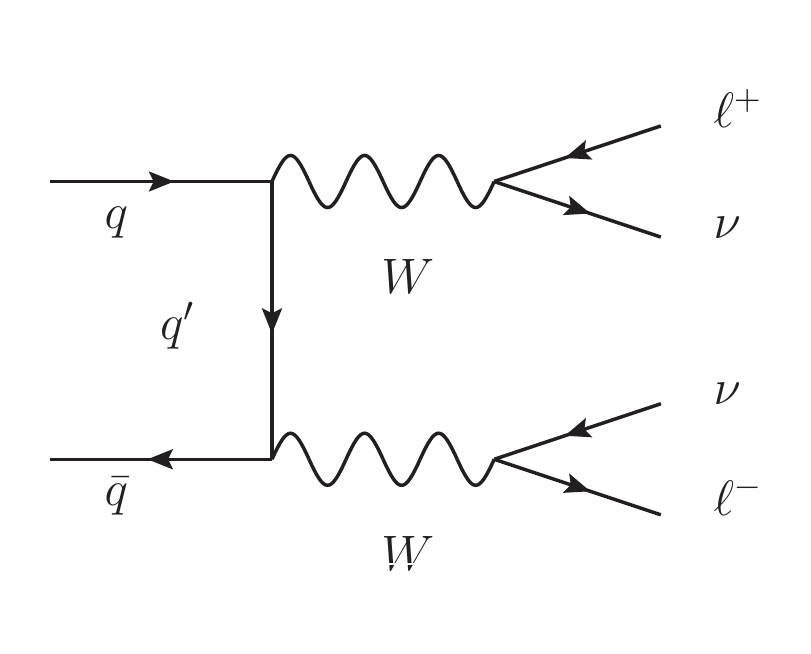}
\label{fig:xx_b}}
\caption{{\bf Diagrams for SUSY benchmark.} Example diagrams describing the signal process involving hypothetical supersymmetric particles $\chi^\pm$ and $\chi^0$ along with charged leptons $\ell^\pm$ and neutrinos $\nu$ (a) and the background process involving $W$ bosons (b). In both cases, the resulting observed particles are two charged leptons, as neutrinos and $\chi^0$ escape undetected.}
\label{fig:xx}
\end{figure}

The signal process is the production of electrically-charged supersymmetric particles $(\chi^\pm)$, which decay to $W$ bosons and an electrically-neutral supersymmetric particle $\chi^0$, which is invisible to the detector. The $W$ bosons decay to charged leptons $\ell$ and invisible neutrinos $\nu$, see Fig.~\ref{fig:xx}.  The final state in the detector is therefore two charged leptons $(\ell\ell)$ and missing momentum carried off by the invisible particles ($\chi^0\chi^0\nu\nu$).  The background process is the production of pairs of $W$ bosons, which  decay to charged leptons $\ell$ and invisible neutrinos $\nu$, see Fig.~\ref{fig:xx}. The visible portion of the signal and background final states both contain two leptons ($\ell\ell$) and large amounts of missing momentum due to the invisible particles.  The classification task requires distinguishing between these two processes using the measurements of the charged lepton momenta and the missing transverse momentum. 

As above, simulated events are generated with the {\sc madgraph}~\cite{madgraph} event
generator assuming 8 TeV collisions of protons as at the latest run of the Large Hadron Collider, with showering and hadronization performed by
{\sc pythia}~\cite{SJO-0601} and detector response simulated by
{\sc delphes}~\cite{delphes}. The masses are set to $m_{\chi^\pm}=200$ GeV and $m_{\chi^0}=100$ GeV.

We focus on the fully leptonic decay mode, in which both $W$ bosons
decay to charged leptons and neutrinos,  $\ell\nu \ell \nu$.  We consider events
which satisfy the requirements:

\begin{itemize}
\item Exactly two electrons or muons, each with $p_{T}>20$ GeV and
  $|\eta|<2.5$; 
\item at least 20 GeV of missing transverse momentum
\end{itemize}

\begin{figure}
\centering
\subfigure[ ]{
\includegraphics[width=1.5in]{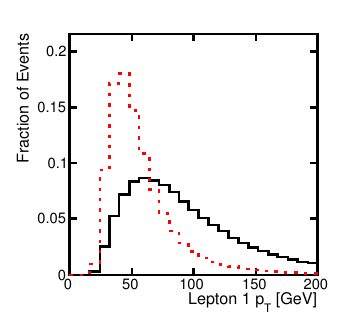}
\label{fig:susy_llvar_a}}
\subfigure[ ]{
\includegraphics[width=1.5in]{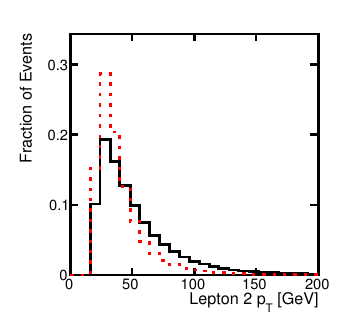}
\label{fig:susy_llvar_b}}
\subfigure[ ]{
\includegraphics[width=1.5in]{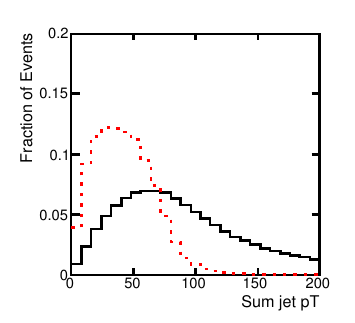}
\label{fig:susy_llvar_c}}
\subfigure[ ]{
\includegraphics[width=1.5in]{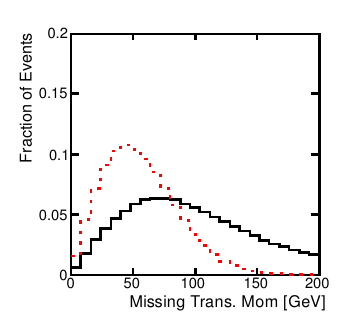}
\label{fig:susy_llvar_d}}
\subfigure[ ]{
\includegraphics[width=1.5in]{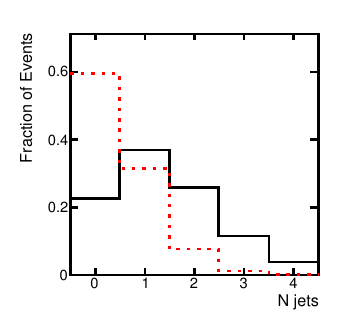}
\label{fig:susy_llvar_e}}
\caption{  {\bf Low-level input features for SUSY benchmark.} Distribution of low-level features in simulated samples for the SUSY signal  (black) and background (red) benchmark processes.}
\label{fig:susy_llvar}
\end{figure}

As above, the basic detector response is used to measure the momentum of each visible particle, in this case the charged leptons. In addition, there may be particle jets induced by radiative processes.  A critical quantity is the missing transverse momentum, $\missET$. Figure~\ref{fig:susy_llvar} gives distributions of low-level features for signal and background processes.

The search for supersymmetric particles is a central piece of the scientific mission of the Large Hadron Collider.  The strategy we applied to the Higgs boson benchmark, of reconstructing the invariant mass of the intermediate state, is not feasible here, as there is too much information carried away by the escaping neutrinos (two neutrinos in this case, compared to one for the Higgs case). Instead,  a great deal of intellectual energy has been spent in attempting to devise features which give additional classification power.  These include high-level features such as:

\begin{itemize}
\item Axial $\missET$: missing transverse energy along the vector defined by the charged leptons,
\item stransverse mass $M_{T2}$: estimating the mass of particles produced in pairs and decaying semi-invisibly~\cite{Barr:2003rg,mt2},
\item $\missET^{Rel}$: $\missET$ if $\Delta\phi\ge \pi/2$, $\missET \sin(\Delta\phi)$ if $\Delta\phi < \pi/2$, where $\Delta\phi$ is the minimum angle between $\missET$ and a jet or lepton,
\item razor quantities $\beta$,$R$, and $M_R$~\cite{razor},
\item super-razor quantities $\beta_{R+1}$, $\cos(\theta_{R+1})$, $\Delta\phi_R^\beta$, $ M_\Delta^R$, $ M_R^T$, and $\sqrt{\hat{s}_R}$  ~\cite{superrazor}.
\end{itemize}

\begin{figure}
\centering
\subfigure[ ]{
\includegraphics[width=1.3in]{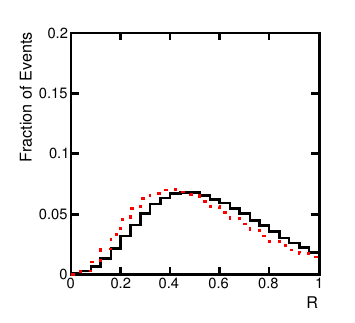}
\label{fig:susy_hlvar_a}}
\subfigure[ ]{
\includegraphics[width=1.3in]{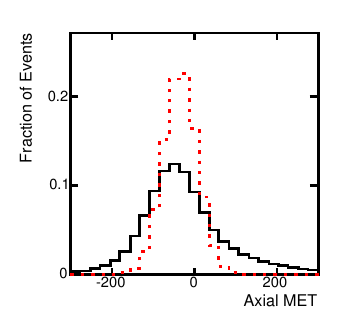}
\label{fig:susy_hlvar_b}}
\subfigure[ ]{
\includegraphics[width=1.3in]{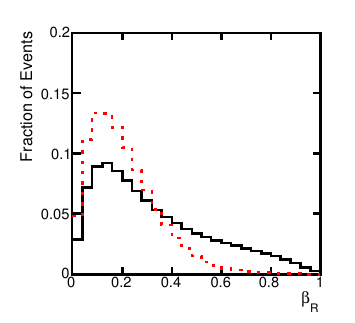}
\label{fig:susy_hlvar_c}}
\subfigure[ ]{
\includegraphics[width=1.3in]{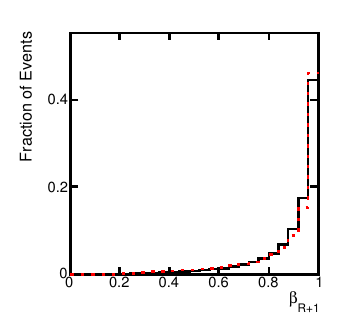}
\label{fig:susy_hlvar_d}}
\subfigure[ ]{
\includegraphics[width=1.3in]{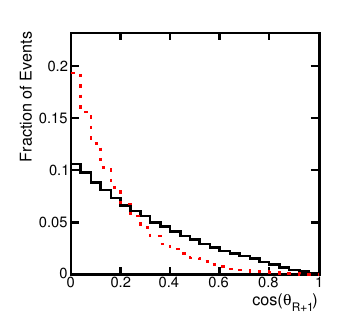}
\label{fig:susy_hlvar_e}}
\subfigure[ ]{
\includegraphics[width=1.3in]{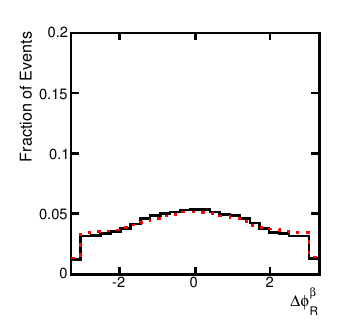}
\label{fig:susy_hlvar_f}}
\subfigure[ ]{
\includegraphics[width=1.3in]{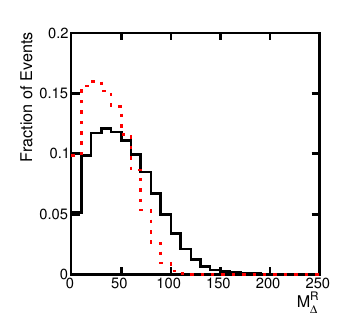}
\label{fig:susy_hlvar_g}}
\subfigure[ ]{
\includegraphics[width=1.3in]{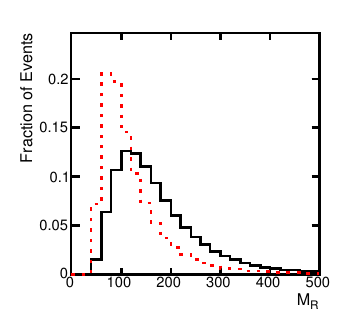}
\label{fig:susy_hlvar_h}}
\subfigure[ ]{
\includegraphics[width=1.3in]{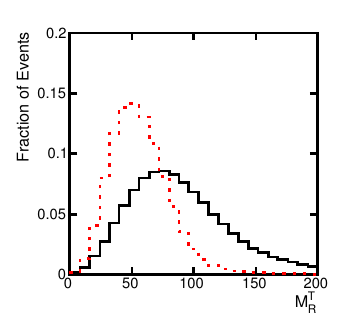}
\label{fig:susy_hlvar_i}}
\subfigure[ ]{
\includegraphics[width=1.3in]{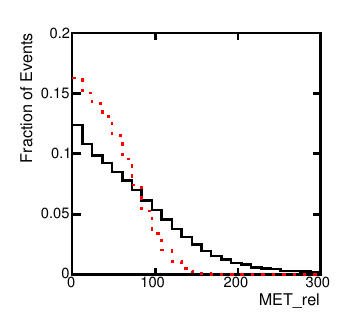}
\label{fig:susy_hlvar_j}}
\subfigure[ ]{
\includegraphics[width=1.3in]{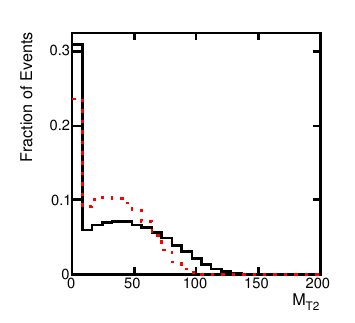}
\label{fig:susy_hlvar_k}}
\subfigure[ ]{
\includegraphics[width=1.3in]{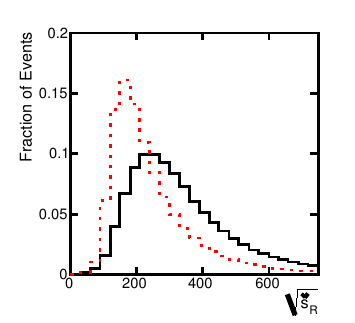}
\label{fig:susy_hlvar_l}}
\caption{ {\bf High-level input features for SUSY benchmark.} Distribution of high-level features in simulated samples for the SUSY signal  (black) and background (red) benchmark processes.}
\label{fig:susy_hlvar}
\end{figure}

See Figure~\ref{fig:susy_hlvar} for distributions of these high-level
features for both signal and background processes.  

A dataset containing five million simulated collision events are available for download at archive.ics.uci.edu/ml/datasets/SUSY.

\subsection*{Current Approach}

Standard techniques in high-energy physics data analyses include feed-forward neural networks with a single hidden layer and boosted-decision trees. We use the widely-used TMVA package~\cite{tmva}, which provides a standardized implementation of common multi-variate learning techniques and an excellent performance baseline.

\subsection*{Deep Learning}

We explored the use of deep neural networks as a practical tool for applications in high-energy physics. Hyper-parameters were chosen using a subset of the HIGGS data consisting of 2.6 million training examples and 100,000 validation examples. Due to computational costs, this optimization was not thorough, but included combinations of the pre-training methods, network architectures, initial learning rates, and regularization methods shown in Supplementary Table~\ref{tab:hyperparameterchoices_deep}. We selected a five-layer neural network with 300 hidden units in each layer, a learning rate of 0.05, and a weight decay coefficient of $1\times 10^{-5}$. Pre-training, extra hidden units, and additional hidden layers significantly increased training time without noticeably increasing performance. To facilitate comparison, shallow neural networks were trained with the same hyper-parameters and the same number of units per hidden layer. Additional training details are provided in the Methods section below.

The hyper-parameter optimization was performed using the full set of HIGGS features. To investigate whether the neural networks were able to learn the discriminative information contained in the high-level features, we trained separate classifiers for each of the three feature sets described above: low-level, high-level, and combined feature sets. For the SUSY benchmark, the networks were trained with the same hyper-parameters chosen for the HIGGS, as the datasets have similar characteristics and the hyper-parameter search is computationally expensive.

\subsection*{Performance}

Classifiers were tested on 500,000 simulated examples generated from the same Monte Carlo procedures as the training sets. We produced Receiver Operating Characteristic (ROC) curves to illustrate the performance of the classifiers. Our primary metric for comparison is the area under the ROC curve (AUC), with larger AUC values indicating higher classification accuracy across a range of threshold choices.

This metric is insightful, as it is directly connected to classification accuracy, which is the quantity optimized for in training.  In practice, physicists may be interested in other metrics, such as signal efficiency at some fixed background rejection, or discovery significance as calculated by $p$-value in the null hypothesis. We choose AUC as it is a standard in machine learning, and is closely correlated with the other metrics. In addition, we calculate discovery significance -- the standard metric in high-energy physics -- to demonstrate that small increases in AUC can represent significant enhancement in discovery significance.  

Note, however, that in some applications the determining factor in the sensitivity to new exotic particles is determined not only by the discriminating power of the selection, but by the uncertainties in the background model itself.  Some portions of the background model may be better understood than others, so that some simulated background collisions have larger associated systematic uncertainties than other collisions.  This can transform the problem into one of reinforcement learning, where per-collision truth labels no longer indicate the ideal network output target. This is beyond the scope of this study, but see Refs.~\cite{whiteson:iaai07,neuro} for stochastic optimizaton strategies for such problems.

\begin{figure}
\centering
\subfigure[ ]{
\includegraphics[width=3in]{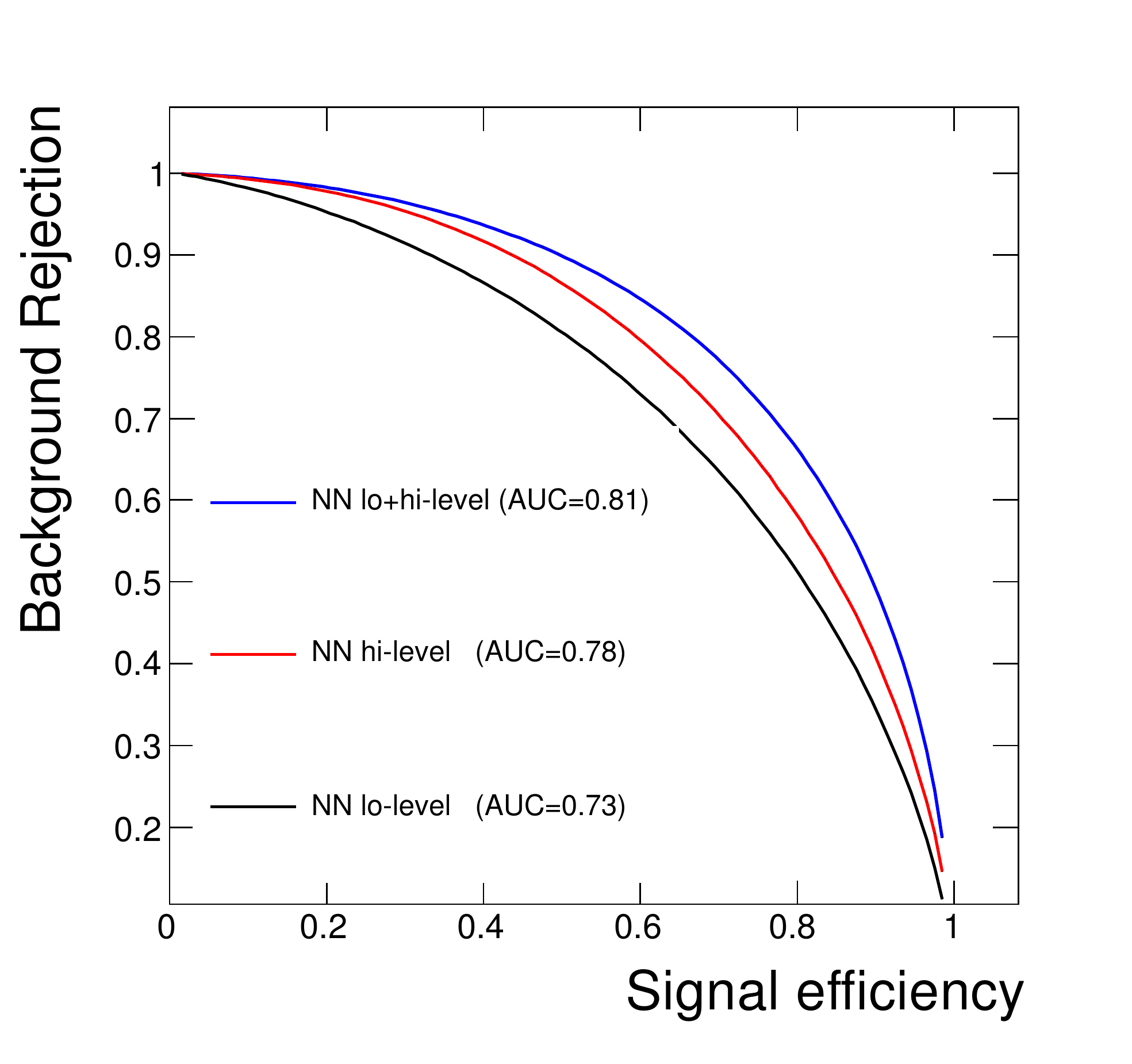}
\label{fig:auc_a}}
\subfigure[ ]{
\includegraphics[width=3in]{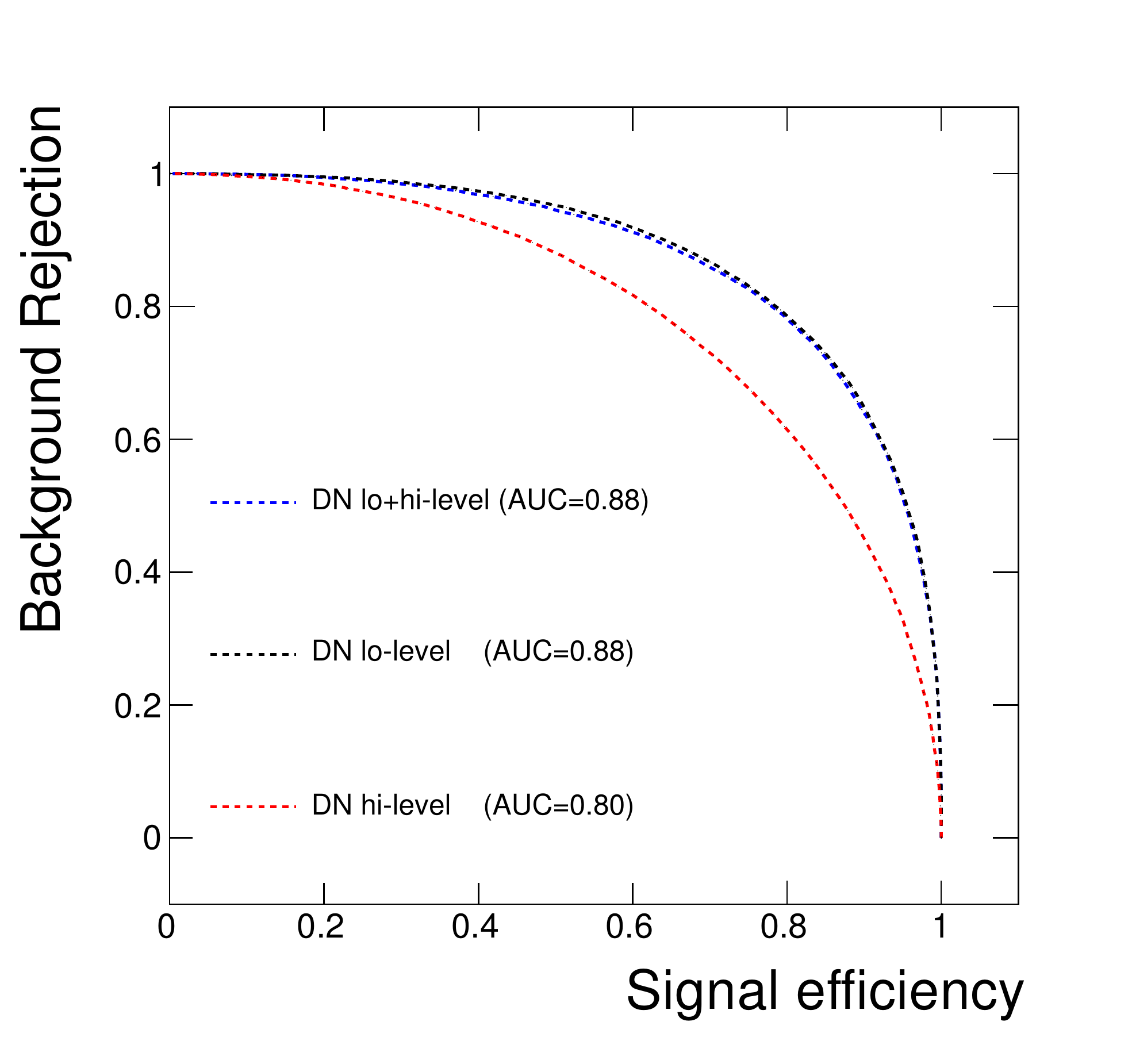}
\label{fig:auc_b}}
\caption{ {\bf Performance for Higgs benchmark.} For the Higgs benchmark, comparison of background rejection versus signal efficiency
  for the traditional learning method (a) and the deep learning method (b) using the low-level features, the
  high-level features and the complete set of features.}
\label{fig:auc}
\end{figure}

Figure~\ref{fig:auc} and  Table~\ref{tab:auc} show the signal efficiency and background rejection for varying thresholds on the output of the neural network (NN) or boosted decision tree (BDT).

\begin{table}
\centering
\caption{ {\bf Performance for Higgs benchmark.} Comparison of the performance of several learning techniques: boosted decision trees (BDT), shallow neural networks (NN), and deep neural networks (DN) for three sets of input features: low-level features, high-level features and the complete set of features. Each neural network was trained five times with different random initializations. The table displays the mean Area Under the Curve (AUC) of the signal-rejection curve in Figure \ref{fig:auc}, with standard deviations in parentheses as well as the expected significance of a discovery (in units of Gaussian $\sigma$) for 100 signal events and $1000\pm50$ background events.}
\label{tab:auc}
\begin{tabular}{llll}
\hline\hline
 & \multicolumn{3}{c}{AUC}\\
Technique & Low-level & High-level & Complete \\
\hline
BDT & 0.73 ($0.01$) & 0.78  ($0.01$)& 0.81  ($0.01$) \\
NN 	& $0.733$ ($0.007$)	& $0.777$ ($0.001$) &  $0.816$ ($0.004$) \\
DN	& $0.880$ ($0.001$)	& $0.800$ ($<0.001$) &  $0.885$ ($0.002$) \\
\hline\hline
 & \multicolumn{3}{c}{Discovery significance}\\
Technique & Low-level & High-level & Complete \\
\hline
NN 	& $2.5\sigma$ & $3.1\sigma$ & $3.7\sigma$\\
DN	& $4.9\sigma$ & $3.6\sigma$ & $5.0\sigma$\\
\hline\hline
\end{tabular}
\end{table}

A shallow NN or BDT trained using only the low-level features performs
significantly worse than one trained with only the high-level
features.  This implies that the shallow NN and BDT are not succeeding in
independently discovering the discriminating power of the high-level
features. This is a well-known problem with shallow learning methods, and motivates the calculation of high-level features.

Methods trained with only the high-level features, however, have a weaker performance than those trained with the full suite of features, which suggests that despite the insight represented by the high-level features, they do not capture all of the information contained in the low-level features. The deep learning techniques show nearly equivalent performance using the low-level features and the complete features, suggesting that they are {\it automatically discovering the insight contained in the high-level features}.  Finally, the deep learning technique finds additional separation power beyond what is contained in the high-level features, demonstrated by the superior performance of the deep network with low-level features to the traditional network using high-level features. These results demonstrate the advantage to using deep learning techniques for this type of problem. 

\begin{figure}
\centering
\subfigure[ ]{
\includegraphics[width=2.5in]{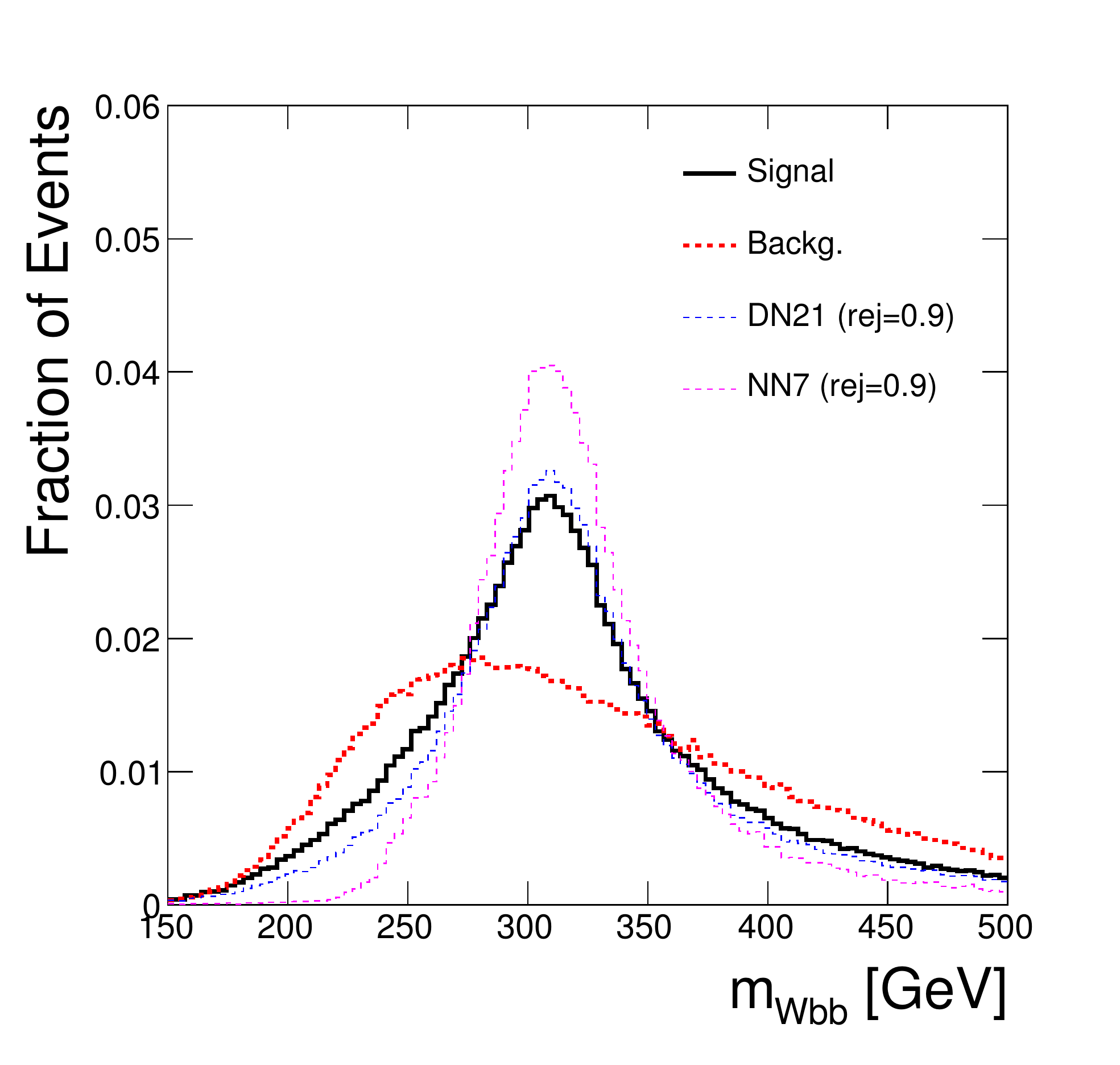}
\label{fig:slice_a}}
\subfigure[ ]{
\includegraphics[width=2.5in]{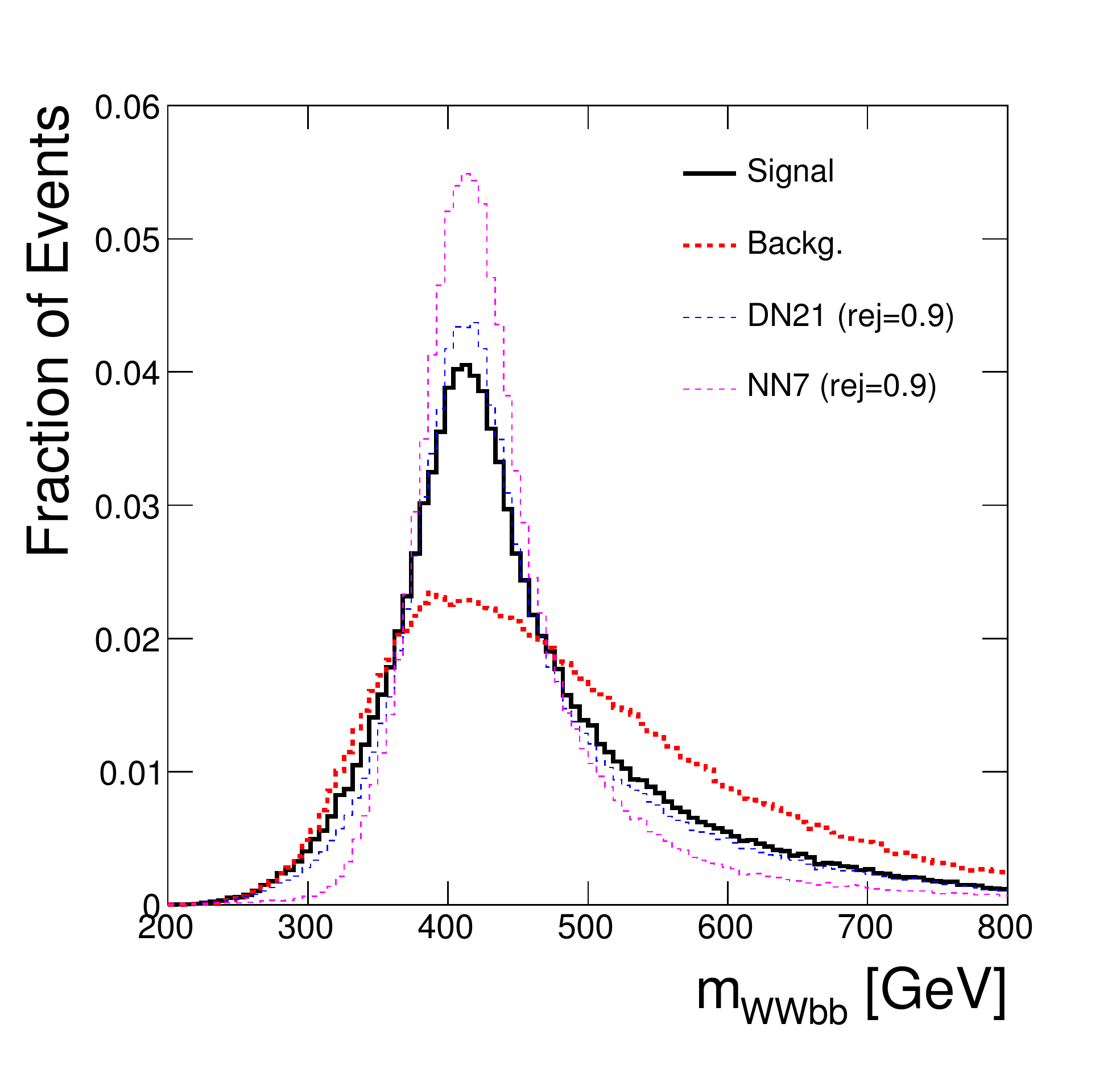}
\label{fig:slice_b}}
\caption{ {\bf Performance comparisons.} Distribution of events for two rescaled input features: (a) $m_{Wbb}$ and (b) $m_{WWbb}$. Shown are pure signal and background distributions, as well as events which pass a threshold requirement which gives a background rejection of 90\% for a deep network with 21 low-level inputs (DN21) and a shallow network with 7 high-level inputs (NN7).} 
\label{fig:slice}
\end{figure}

The internal representation of a NN is notoriously difficult to reverse engineer. To gain some insight into the mechanism by which the deep network (DN) is improving upon the discrimination in the high-level physics features, we compare the distribution of simulated events selected by a minimum threshold on the NN or DN output, chosen to give equivalent rejection of 90\% of the background events.  Figure~\ref{fig:slice} shows events selected by such thresholds in an mixture of 50\% signal and 50\% background collisions, compared to pure distributions of signal and background.  The NN preferentially selects events with values of the features close to the characteristic signal values and away from background-dominated values.  The DN, which has a higher efficiency for the equivalent rejection, selects events near the same signal values, but {\it also retains events away from the signal-dominated region}. The likely explanation is that the DN has discovered the same signal-rich region identified by the physics features, but has in addition found avenues to carve into the background-dominated region.

In the case of the SUSY benchmark, the deep neural networks again perform better than the shallow networks. The improvement is less dramatic, though statistically significant.

 An additional boost in performance is obtained by using the dropout training algorithm, in which we stochastically drop neurons in the top hidden layer with $50\%$ probability during training. For deep networks trained with dropout, we achieve an AUC of 0.88 on both the low-level and complete feature sets. Table \ref{tab:auc_susy}, Supplementary Fig.~\ref{fig:susy_auc} and Supplementary Fig.~\ref{fig:susy_auc2} compare the performance of shallow and deep networks for each of the three sets of input features.

  In this SUSY case, neither the high-level features nor the deep network finds dramatic gains over the shallow network of low-level features. The power of the deep network to automatically find non-linear features reveals something about the nature of the classification problem in this case: it suggests that there may be little gain from further attempts to manually construct high-level features.  

\begin{table}
\centering
\caption{{\bf Performance comparison for the SUSY benchmark.} Each model was trained five times with different weight initializations. The mean AUC is shown with standard deviations in parentheses as well as the expected significance of a discovery (in units of Gaussian $\sigma$) for 100 signal events and $1000\pm50$ background events.}
\label{tab:auc_susy}
\begin{tabular}{llll}
\hline\hline
& \multicolumn{3}{c}{AUC} \\
Technique & Low-level & High-level & Complete \\
\hline
BDT						 & $0.850$  $(0.003)$ &  $0.835$ $(0.003)$ &  $0.863$ $(0.003)$ \\
NN 					 	 & $0.867$ ($0.002$) &  $0.863$ ($0.001$) &  0.875 ($<0.001$) \\
NN$_{dropout}$ 			 & $0.856$ ($<0.001$) & 0.859 ($<0.001$) & 0.873 ($<0.001$) \\
DN 		 	 & $0.872$ ($0.001$) & 0.865 ($0.001$) & 0.876 ($<0.001$) \\
DN$_{dropout}$ & $0.876$ ($<0.001$) & 0.869 ($<0.001$) & 0.879 ($<0.001$) \\
\hline\hline
 & \multicolumn{3}{c}{Discovery significance}\\
Technique & Low-level & High-level & Complete \\
\hline
NN 	& $6.5\sigma$ & $6.2\sigma$ & $6.9\sigma$\\
DN	& $7.5\sigma$ & $7.3\sigma$ & $7.6\sigma$\\
\hline\hline
\end{tabular}
\end{table}

\subsection*{Analysis}

To highlight the advantage of deep networks over shallow networks with a similar number of parameters, we performed a thorough hyper-parameter optimization for the class of single-layer neural networks over the hyper-parameters specified in Supplementary Table \ref{tab:hyperparameterchoices_shallow} on the HIGGS benchmark. The largest shallow network had $300,001$ parameters, slightly more than the $279,901$ parameters in the largest deep network, but these additional hidden units did very little to increase performance over a shallow network with only $30,001$ parameters. Supplementary Table \ref{tab:auc_higgs_layers} compares the performance of the best shallow networks of each size with deep networks of varying depth.

While the primary advantage of deep neural networks is their ability to \emph{automatically} learn high-level features from the data, one can imagine facilitating this process by pre-training a neural network to compute a particular set of high-level features. As a proof of concept, we demonstrate how deep neural networks can be trained to compute the high-level HIGGS features with a high degree of accuracy (Supplementary Table \ref{tab:regression}). Note that such a network could be used as a module within a larger neural network classifier.

\section{Discussion}

It is widely accepted in experimental high-energy physics that machine learning techniques can provide powerful boosts to searches for exotic particles.  Until now, physicists have reluctantly accepted the limitations of the shallow networks employed to date; in an attempt to circumvent these limitations, physicists manually construct helpful non-linear feature combinations to guide the shallow networks.

Our analysis shows that recent advances in deep learning techniques may lift these limitations by {\bf automatically discovering powerful non-linear feature combinations and providing better discrimination power than current classifiers} -- even when aided by manually-constructed features.  This appears to be the first such demonstration in a semi-realistic case.

We suspect that the novel environment of high energy physics, with high volumes of relatively low-dimensional data containing rare signals hiding under enormous backgrounds, can inspire new developments in machine learning tools.  Beyond these simple benchmarks, deep learning methods may be able to tackle thornier problems with multiple backgrounds, or lower-level tasks such as identifying the decay products from the high-dimensional raw detector output.

\section{Methods}

\subsection*{Neural Network Training}

In training the neural networks, the following hyper-parameters were predetermined without optimization. Hidden units all used the $\emph{tanh}$ activation function. Weights were initialized from a normal distribution with zero mean and standard deviation $0.1$ in the first layer, $0.001$ in the output layer, and  $0.05$ all other hidden layers. Gradient computations were made on mini-batches of size $100$. A momentum term increased linearly over the first $200$ epochs from $0.9$ to $0.99$, at which point it remained constant. The learning rate decayed by a factor of $1.0000002$ every batch update until it reached a minimum of $10^{-6}$. Training ended when the momentum had reached its maximum value and the minimum error on the validation set (500,000 examples) had not decreased by more than a factor of $0.00001$ over 10 epochs. This early stopping prevented overfitting and resulted in each neural network being trained for 200-1000 epochs. 

Autoencoder pretraining was performed by training a stack of single-hidden-layer autoencoder networks as in \cite{bengio_greedy_2007}, then fine-tuning the full network using the class labels. Each autoencoder in the stack used tanh hidden units and linear outputs, and was trained with the same initialization scheme, learning algorithm, and stopping parameters as in the fine-tuning stage. When training with dropout, we increased the learning rate decay factor to $1.0000003$, and only ended training when the momentum had reached its maximum value and the error on the validation set had not decreased for 40 epochs.

\subsection*{Datasets}

The data sets were nearly balanced, with $53\%$ positive examples in the HIGGS data set, and $46\%$ positive examples in the SUSY data set. Input features were standardized over the entire train/test set with mean zero and standard deviation one, except for those features with values strictly greater than zero -- these we scaled so that the mean value was one.

\subsection*{Computation}

Computations were performed using machines with 16 Intel Xeon cores, an NVIDIA Tesla C2070 graphics processor, and 64 GB memory. All neural networks were trained using the GPU-accelerated Theano and Pylearn2 software libraries \cite{bergstra_theano:_2010,pylearn2_arxiv_2013}. Our code is available at https://github.com/uci-igb/higgs-susy.

\section{Acknowledgments} 
We are grateful to Kyle Cranmer, Chris Hays, Chase Shimmin, Davide Gerbaudo, Bo Jayatilaka, Jahred Adelman, and Shimon Whiteson for their insightful comments. We wish to acknowledge a hardware grant from NVIDIA.

\section{Author Contributions}

PB conceived the idea of applying deep learning methods to high energy physics and particle detection. DW chose the benchmarks and generated the data and structured the problem. PB and PS designed the architectures and the algorithms. PS implemented the code and performed the experiments. All authors analyzed the results and contributed to writing up the manuscript.

\section{References}
\bibliographystyle{naturemag}
\bibliography{paper,2013DeepLearningPhysics,physics,baldi,nn} 

\clearpage

\setcounter{figure}{0} \renewcommand{\thefigure}{\arabic{figure}}
\setcounter{table}{0} \renewcommand{\thetable}{\arabic{table}}

\begin{figure}
\centering
\subfigure[ ]{
\includegraphics[width=2in]{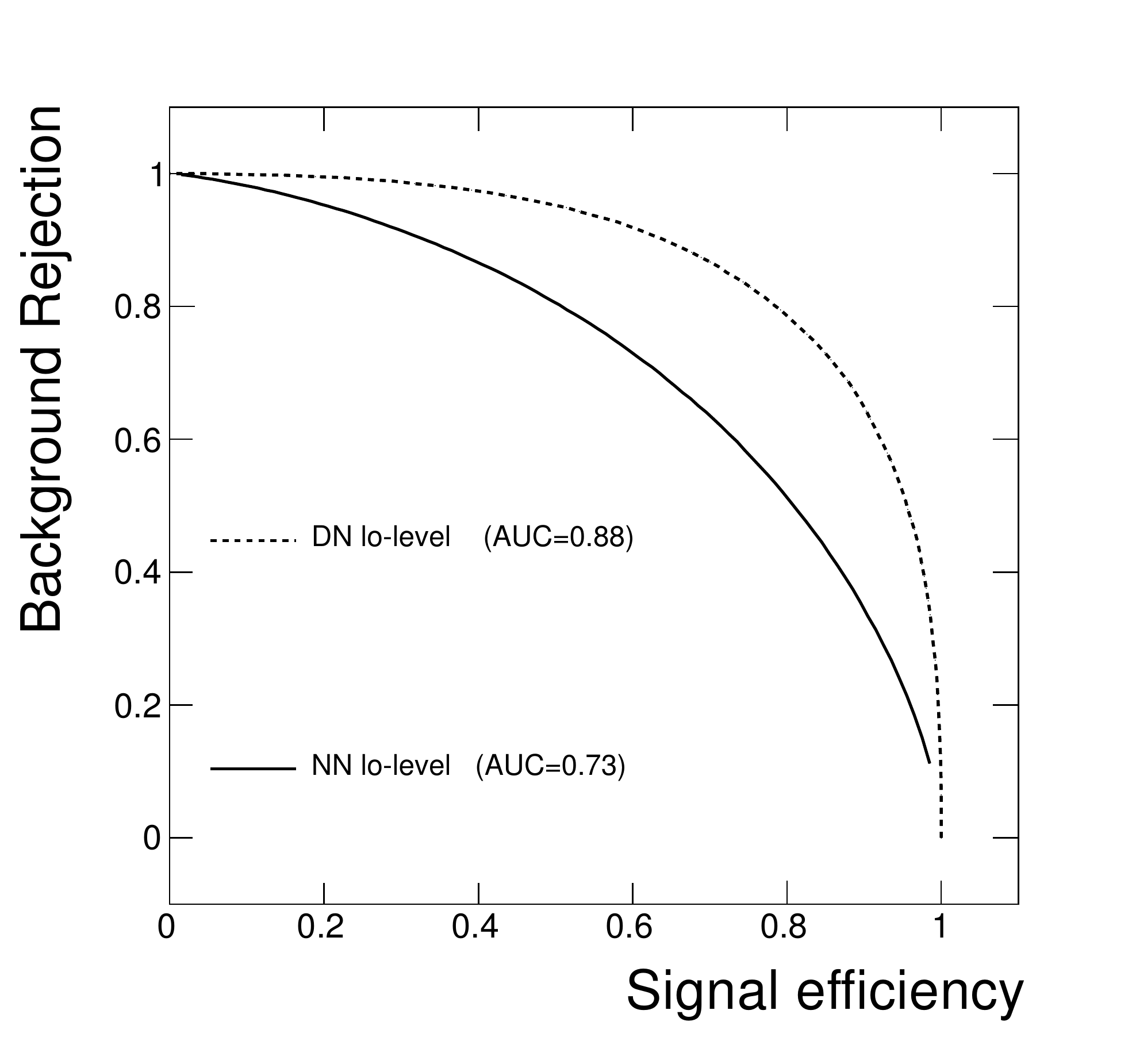}
\label{fig:auc2_a}}
\subfigure[ ]{
\includegraphics[width=2in]{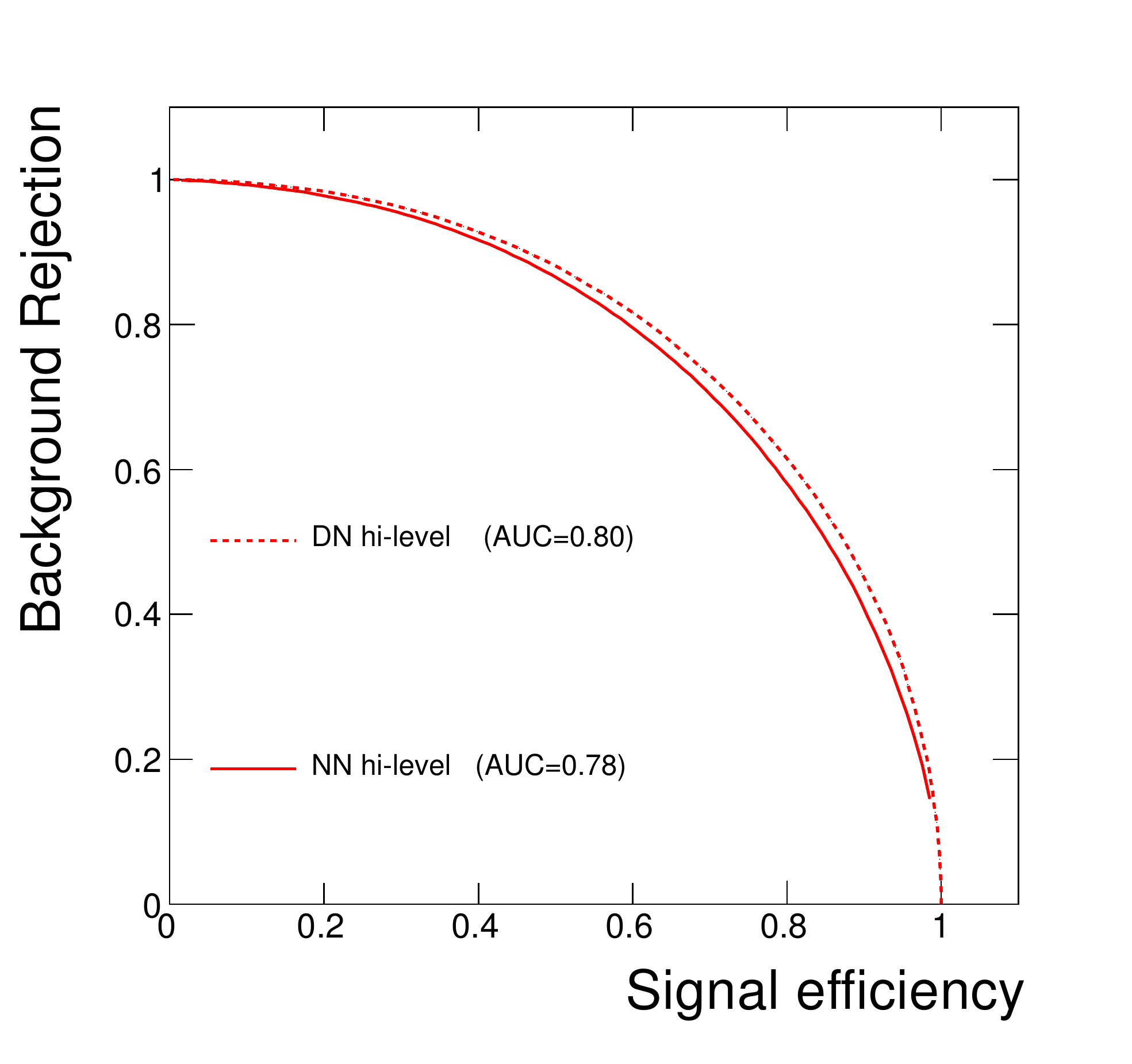}
\label{fig:auc2_b}}
\subfigure[ ]{
\includegraphics[width=2in]{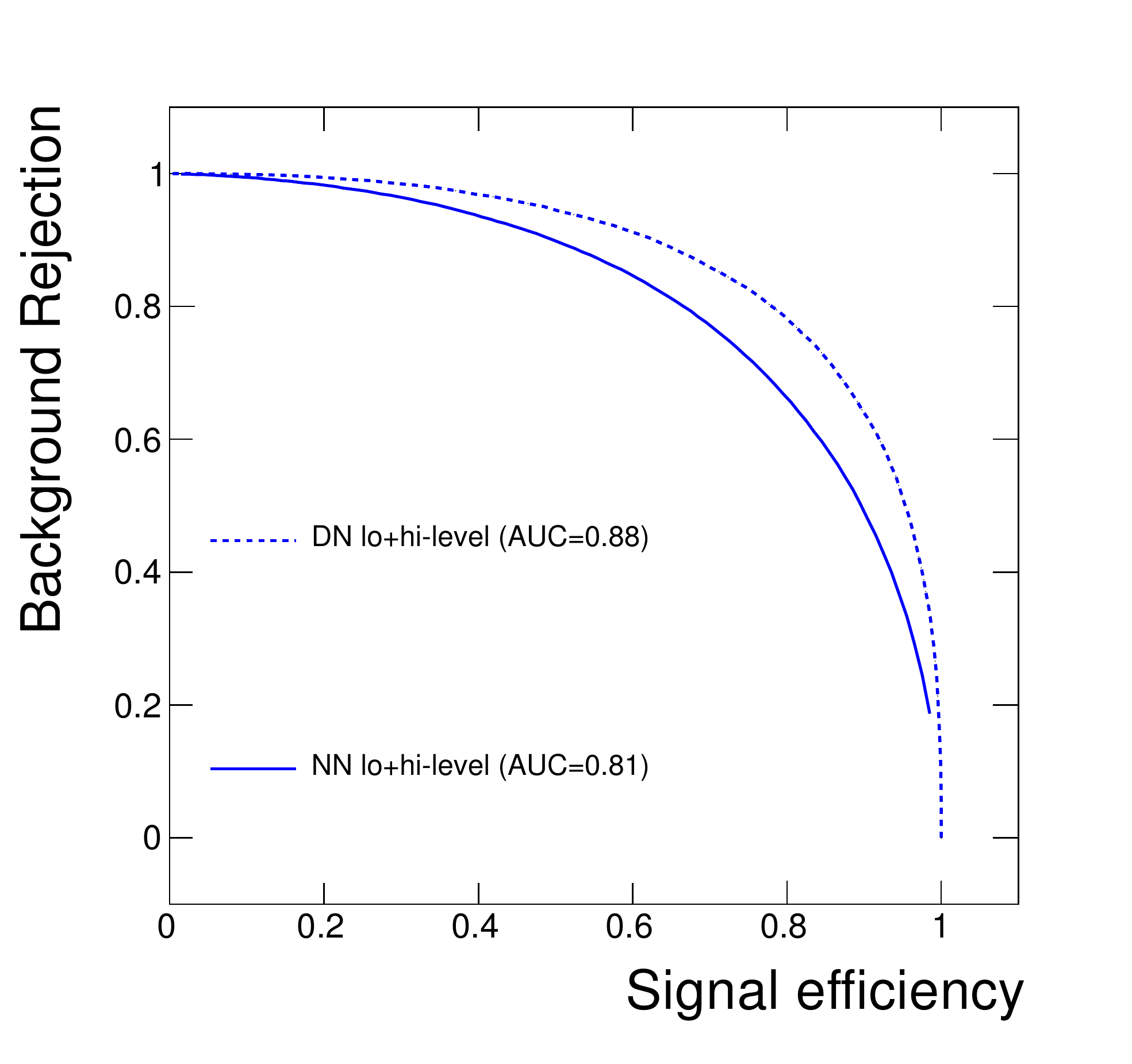}
\label{fig:auc2_c}}
\renewcommand{\figurename}{Supplementary Figure}
\caption{{\bf Details of performance for Higgs benchmark.} For the HIGGS benchmark, comparison of background rejection versus signal efficiency for the low-level features (a), high-level features (b) and complete set of features (c) using traditional and deep learning methods.}
\label{fig:auc2}
\end{figure}

\begin{figure}
\centering
\subfigure[ ]{
\includegraphics[width=3in]{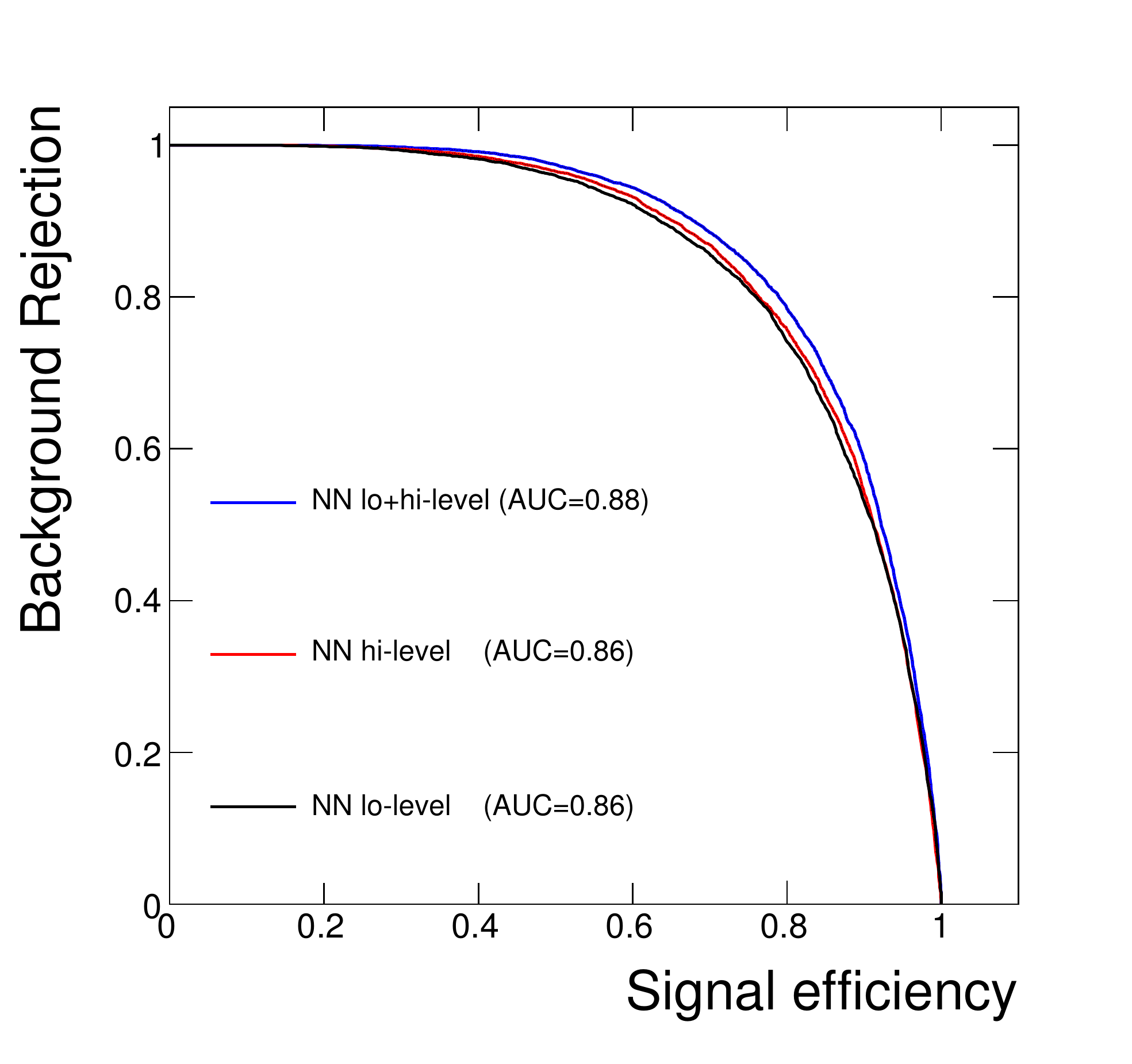}
\label{fig:susy_auc_a}}
\subfigure[ ]{
\includegraphics[width=3in]{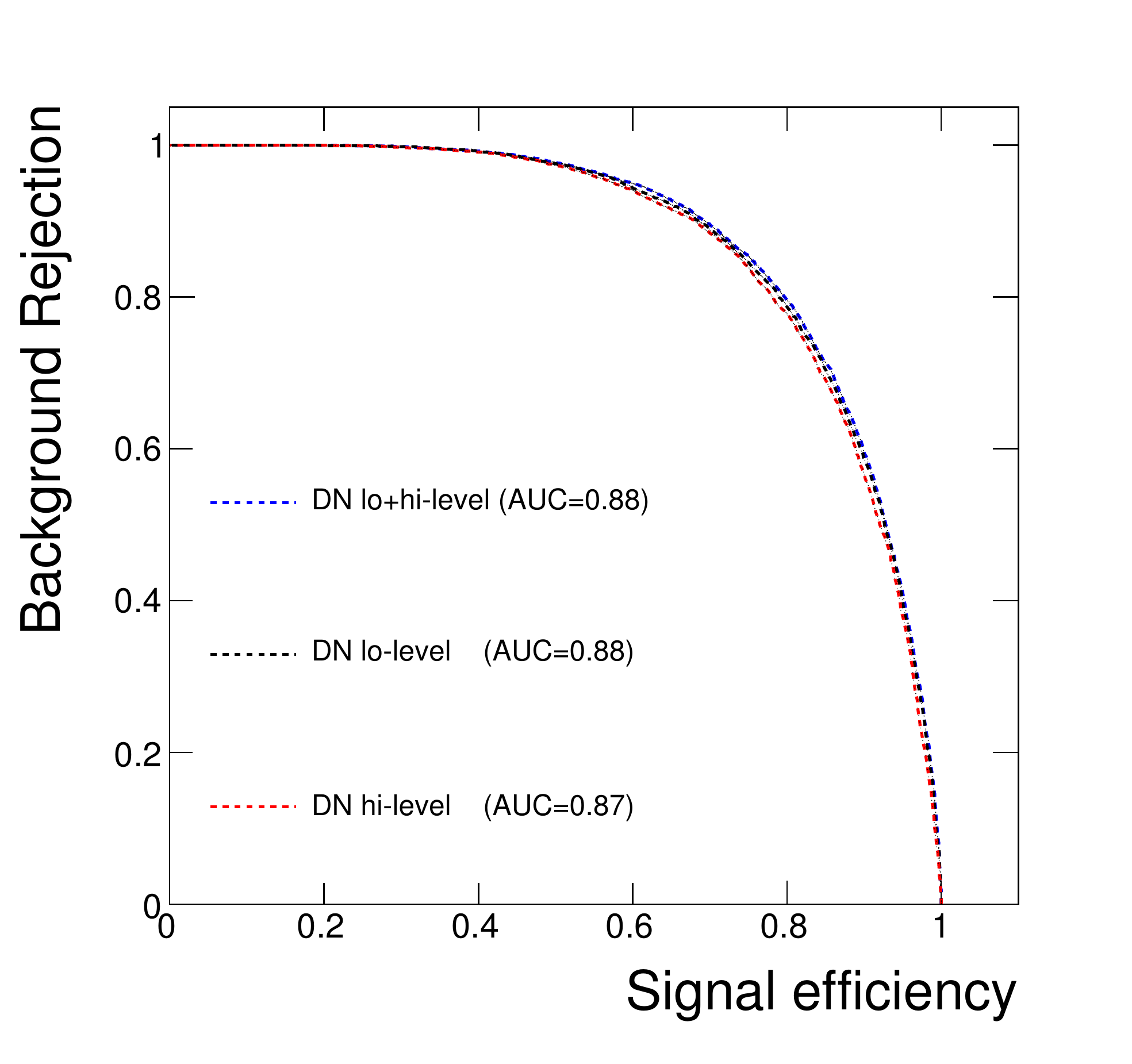}
\label{fig:susy_auc_b}}
\renewcommand{\figurename}{Supplementary Figure}
\caption{ {\bf Performance for SUSY benchmark.} Comparison of background rejection versus signal efficiency
  for the traditional learning method (a) and the deep learning method (b) using the low-level features, the
  high-level features and the complete set of features.}
\label{fig:susy_auc}
\end{figure}

\begin{figure}
\centering
\subfigure[ ]{
\includegraphics[width=2in]{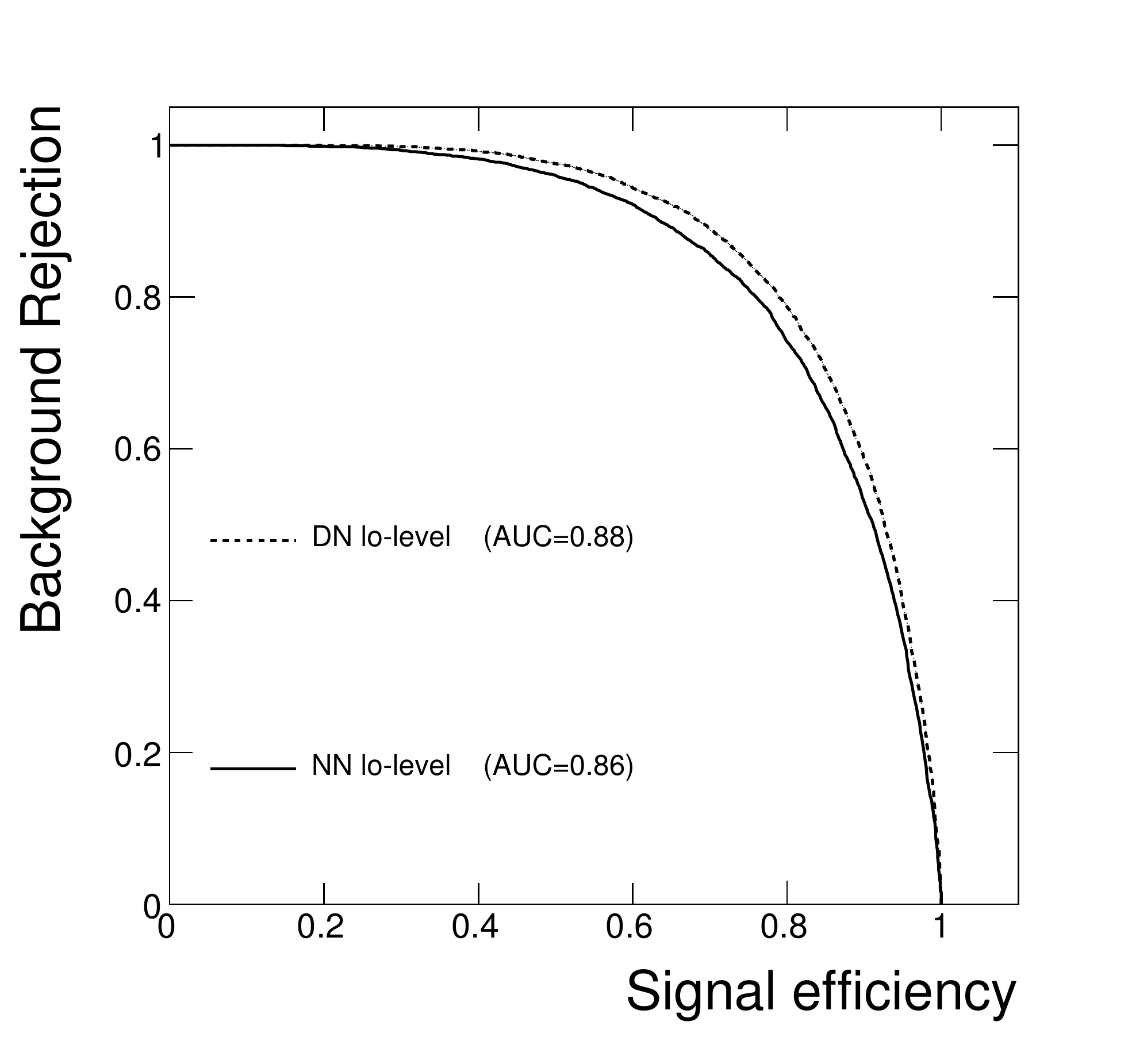}
\label{fig:susy_auc2_a}}
\subfigure[ ]{
\includegraphics[width=2in]{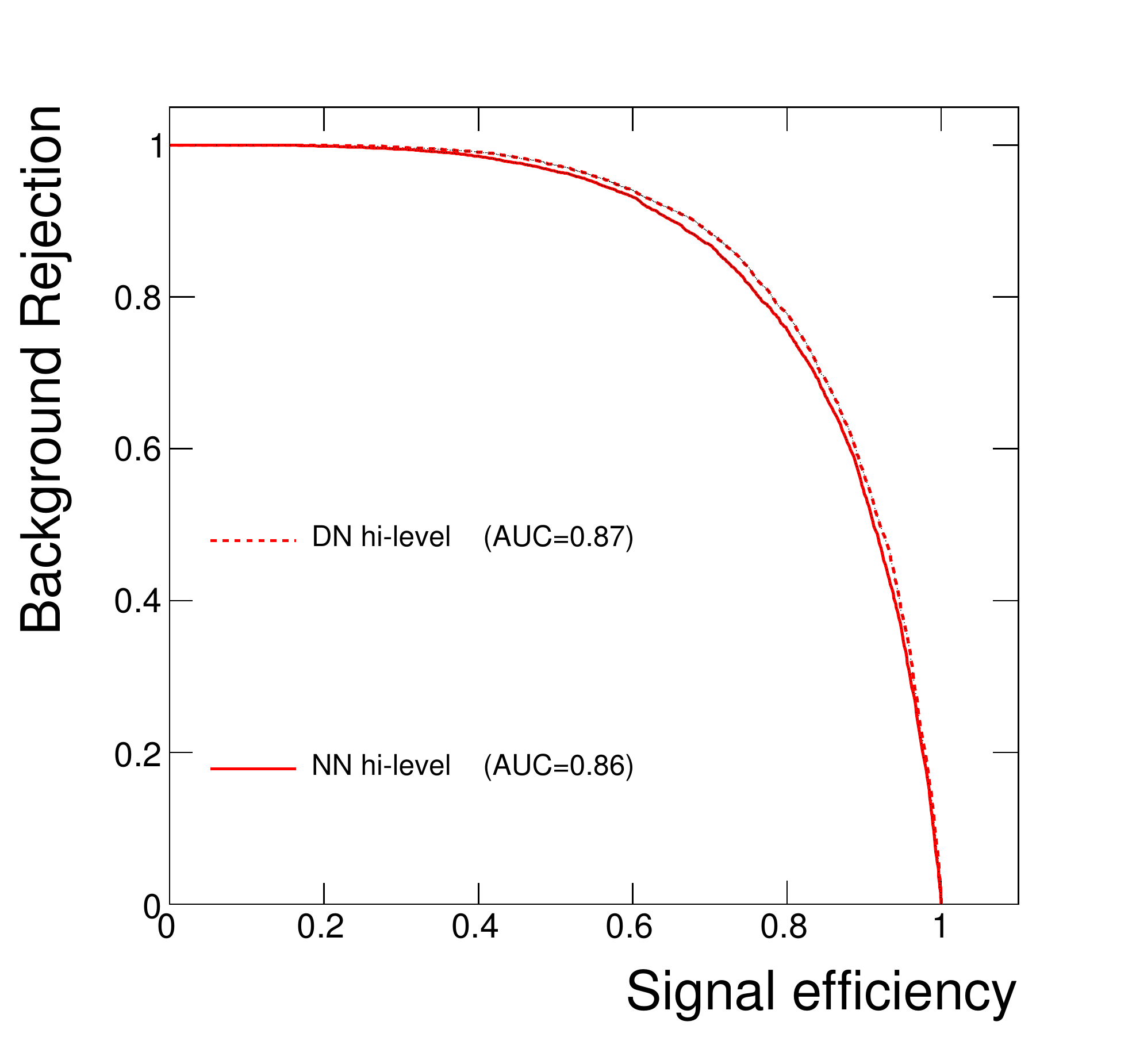}
\label{fig:susy_auc2_b}}
\subfigure[ ]{
\includegraphics[width=2in]{fig11b.pdf}
\label{fig:susy_auc2_c}}
\renewcommand{\figurename}{Supplementary Figure}
\caption{ {\bf Details of performance for SUSY benchmark.} Comparison of background rejection versus signal efficiency
  for the low-level features (a), high-level features (b) and complete set of features (c) using traditional and deep learning methods.}
\label{fig:susy_auc2}
\end{figure}

\clearpage

\begin{table}
\centering
\renewcommand{\tablename}{Supplementary Table}
\caption{{\bf Hyper-parameter choices for deep networks.} Shown are the considered value for each hyper-parameter.}
\label{tab:hyperparameterchoices_deep}
\begin{tabular}{lr}
\hline\hline
Hyper-parameters & Choices \\
\hline
Depth & 2, 3, 4, 5, 6 layers \\
Hidden units per layer & 100, 200, 300, 500 \\
Initial learning rate & 0.01, 0.05  \\
Weight decay & 0, $10^{-5}$  \\
Pre-training & none, autoencoder\\
\hline\hline
\end{tabular}
\end{table}

\begin{table}
\centering
\renewcommand{\tablename}{Supplementary Table}
\caption{{\bf Hyper-parameter choices for shallow networks.} Shown are the considered value for each hyper-parameter.}
\label{tab:hyperparameterchoices_shallow}
\begin{tabular}{lr}
\hline\hline
Hyper-parameters & Choices \\
\hline
Hidden units & 300, 1000, 2000, 10000 \\
Initial learning rate & 0.05, 0.005, 0.0005 \\
Weight decay & 0, $10^{-5}$  \\
\hline\hline
\end{tabular}
\end{table}

\begin{table}
\renewcommand{\tablename}{Supplementary Table}
\caption{{\bf Study of network size and depth.} Comparison of shallow networks with different numbers of hidden units (single hidden layer), and deep networks with varying hidden layers in terms of the Area Under the ROC Curve (AUC) for the HIGGS benchmark. The deep networks have 300 units in each hidden layer.}
\label{tab:auc_higgs_layers}
\begin{tabular}{llll}
\hline\hline
 & \multicolumn{3}{c}{AUC}\\
Technique & Low-level & High-level & Complete \\
\hline
NN 300-hidden	& $0.733$ & $0.777$ &  $0.816$ \\
NN 1000-hidden 	& $0.788$  & $0.783$  &  $0.841$  \\
NN 2000-hidden 	& $0.787$ & $0.788$  &  $0.842$ \\
NN 10000-hidden 	& $0.790$ & $0.789$  &  $0.841$ \\
DN 3 layers 	& $0.836$ & $0.791$  &  $0.850$ \\
DN 4 layers 	& $0.868$ & $0.797$  &  $0.872$ \\
DN 5 layers & $0.880$ & $0.800$  &  $0.885$  \\
DN 6 layers 	& $0.888$ & $0.799$  &  $0.893$ \\
\hline\hline
\end{tabular}
\end{table}

\begin{table}
\renewcommand{\tablename}{Supplementary Table}
\caption{ {\bf Error Analysis. } Mean Squared Error (MSE) of networks trained to compute the seven high-level features from the 21 low-level features on HIGGS. All networks were trained using the same hyperparameters as the deep networks in the previous experiments, except with linear output units, no momentum, and a smaller initial learning rate of $0.005$.}
\label{tab:regression}
\begin{tabular}{lc}
\hline\hline
Technique & Feature Regression MSE \\
\hline
Linear Regression & $0.1468$ \\
NN &  $0.0885$ \\
DN 3 layers &  $0.0821$ \\
DN 4 layers &  $0.0818$ \\
DN 5 layers &  $0.0815$ \\
DN 6 layers &  $0.0812$ \\
\hline\hline
\end{tabular}
\end{table}

\end{document}